ARTICLE  OPEN

# Reversible magnetic switching of high-spin molecules on a giant Rashba surface

Jens Kügel [1], Michael Karolak[2], Andreas Krönlein[1], David Serrate[3,4], Matthias Bode [1,5] and Giorgio Sangiovanni[2]

The quantum mechanical screening of a spin via conduction electrons depends sensitively on the environment seen by the magnetic impurity. A high degree of responsiveness can be obtained with metal complexes, as the embedding of a metal ion into an organic molecule prevents intercalation or alloying and allows for a good control by an appropriate choice of the ligands. There are therefore hopes to reach an "on demand" control of the spin state of single molecules adsorbed on substrates. Hitherto one route was to rely on "switchable" molecules with intrinsic bistabilities triggered by external stimuli, such as temperature or light, or on the controlled dosing of chemicals to form reversible bonds. However, these methods constrain the functionality to switchable molecules or depend on access to atoms or molecules. Here, we present a way to induce bistability also in a planar molecule by making use of the environment. We found that the particular "habitat" offered by an antiphase boundary of the Rashba system $BiAg_2$ stabilizes a second structure for manganese phthalocyanine molecules, in which the central Mn ion moves out of the molecular plane. This corresponds to the formation of a large magnetic moment and a concomitant change of the ground state with respect to the conventional adsorption site. The reversible spin switch found here shows how we can not only rearrange electronic levels or lift orbital degeneracies via the substrate, but even sway the effects of many-body interactions in single molecules by acting on their surrounding.

## INTRODUCTION

The control of the charge- and spin state of single molecules is key not only for a fundamental understanding of the underlying interactions but also for the design of molecular electronic and spintronic devices. In this respect, phthalocyanine (Pc) molecules have become particularly popular due to their chemical and thermal stability and the possibility to tune the electronic and magnetic properties by the choice of the central metal ion. Beyond these molecule-specific properties, also reversible switching of the charge- and/or of the spin state of single surface-mounted phthalocyanine molecules was demonstrated, which is predominantly based on one of the following two principles. On the one hand, the adsorption of atoms[1] or small molecules[2] into the molecular frame, more precisely on the metal ion of the molecule, and their desorption from the molecules can be used to redistribute the $d$-shell electrons of the metal ion, thereby potentially changing the spin state of the molecule. The adsorption of these atoms or small molecules is achieved by dosing a small amount of them, whereas the desorption can accomplished either by the application of a voltage pulse with a STM tip over the molecule[1,2] or by increasing the temperature.[1] On the other hand, phthalocyanine molecules that are switchable between different configurations are used,[3–5] leading to different electronic and/or magnetic properties of the molecules. It was demonstrated that this switching can be achieved by means of bias voltages[5] or temperature.[4] The methods described before either require the insertion of gases or are restricted to non-planar phthalocyanine molecules which themselves can be stable in more than one configuration.

Here, we present a different way of designing a reversible molecular switch based on a MnPc molecule. This molecule is well known for its relatively high magnetic moment leading to the Kondo effect—the screening of the magnetic moment by itinerant electrons of the substrate—on various surfaces.[2,6–11] However, its planar structure usually allows only one stable configuration, thereby limiting its suitability for switching at first glance. We circumvent this issue, by choosing a proper environment for the MnPc molecule, which enables us to control the position of the central metal ion with the electric field of an STM tip and thereby the magnetic and electronic properties of the molecule. In detail, we positioned a MnPc on top of an antiphase boundary of the $\sqrt{3} \times \sqrt{3}$ reconstructed surface alloy of bismuth on Ag(111) ($BiAg_2$).[12] By using the electric field of the STM tip, this specific MnPc molecule can be switched between two different states, which can be identified by their topographic appearance (protrusion vs. depression in the center of the molecule) and their spectroscopic fingerprint around the Fermi energy (resonances vs. no features).

By means of DFT + U calculations we show that the unique environment with the MnPc molecule bridging an antiphase boundary plays a crucial role. Due to bonding to the edges of the antiphase boundary the central metal ion can not only be stabilized within the plane of the molecular frame, as common to the class of metal phthalocyanines, but also outside the molecular plane, thereby leading to a second stable configuration with the Mn ion positioned between the molecular frame and the substrate. This structural bistability is accompanied by a strong variation of the magnetic moment, which increases from $2.55\mu_B$ to $4.85\mu_B$ as the central metal ion is transferred out of the molecular plane.

[1]Physikalisches Institut, Experimentelle Physik II, Universität Würzburg, Am Hubland, 97074 Würzburg, Germany; [2]Institut für Theoretische Physik und Astrophysik, Universität Würzburg, Am Hubland, 97074 Würzburg, Germany; [3]Instituto de Nanociencia de Aragón & Laboratorio de Microscopias Avanzadas, University of Zaragoza, E-50018 Zaragoza, Spain; [4]Departamento Física Materia Condensada, University of Zaragoza, E-50018 Zaragoza, Spain and [5]Wilhelm Conrad Röntgen-Center for Complex Material Systems (RCCM), Universität Würzburg, Am Hubland, D-97074 Würzburg, Germany
Correspondence: Jens Kügel (jens.kuegel@physik.uni-wuerzburg.de) or Giorgio Sangiovanni (sangiovanni@physik.uni-wuerzburg.de)

## RESULTS AND DISCUSSION

### Topography and spectroscopy of molecular switch

In Fig. 1a a topographic image of three MnPc molecules adsorbed on the giant Rashba surface $BiAg_2$—the $\sqrt{3}\times\sqrt{3}R30$ reconstruction of bismuth on Ag(111)—is shown. The arms of the molecules are rotated by 45° with respect to the high symmetry directions $\langle 2\bar{1}\bar{1}\rangle$ and $\langle 01\bar{1}\rangle$ of the Ag(111) surface. Similar to the adsorption on other substrates, the MnPc exhibits a central protrusion. All molecules shown in Fig. 1a are adsorbed in close vicinity to an antiphase boundary (dark line), which generally form during the growth process if two areas of the $BiAg_2$ surface alloy are separated by an additional row of silver atoms hindering the formation of an uniform phase.

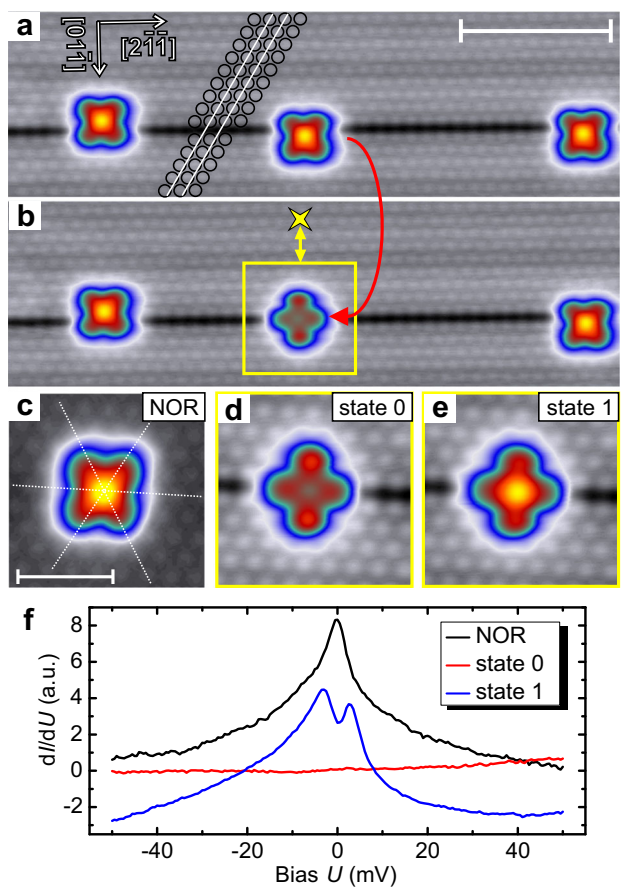

**Fig. 1** **a** Topography of MnPc on $BiAg_2$ showing three molecules in close vicinity to an antiphase boundary (black horizontal line). The phase shift between the two sides of the antiphase boundary is indicated by white lines, which are positioned on top of the Bi atoms in the upper patch (marked by black circles) and shifted in the lower patch. Scale bar: 5 nm. **b** Topography of the same area as in **a** after moving the central MnPc molecule onto the antiphase boundary (red arrow). **c** MnPc on flat $BiAg_2$ ("NOR") adsorbs with the central ion being positioned over a Bi atom (cf. white lines). Scale bar: 2 nm. **d** Topography of MnPc molecule in the deactivated state (state 0) on the antiphase boundary. **e** Topography after activation of the molecule [scan parameters **a**–**d**: $U = -0.1$ V, $I = 0.1$ nA]. **f** Spectroscopy curves taken at the central ion of the MnPc molecules in the NOR position (black curve) or located on the antiphase boundary, while being in the deactivated (red curve) or activated (blue curve) state. A spectrum of the substrate has been subtracted from all three to reduce the effect of tip features. (parameters: $T \approx 4.5$ K, $U_{mod} = 1$ mV, $I = 0.2$ nA)

By applying a bias voltage of $U = 1.5$ V and a tunneling current of a few nA a MnPc molecule can be moved into the antiphase boundary, as indicated by the red arrow in Fig. 1b. After this manipulation, the central ion of the molecule appears as a depression and the arms are oriented along the high symmetry directions of the substrate with two of them bridging and the other two lying along the antiphase boundary. We refer to this in the following as "state 0", a close-up of which can be seen in Fig. 1d. The differences with respect to the configuration that we call "NOR" can be appreciated by comparison with Fig. 1c, which shows MnPc on flat $BiAg_2$ ("normal" position). When the MnPc molecule is on the antiphase boundary, we can induce the switch via the following procedure: starting from "state 0", we move the tip away from the molecule [e.g., to the position marked by a yellow cross in Fig. 1b] and applying a voltage pulse of $U_{Pulse} < -0.8V$. With a certain probability, which will be discussed later on, the bias pulse changes the topography of the molecule, as can be seen in Fig. 1e, where the center of the MnPc molecule appears as a protrusion. This activated state of the molecule will be labeled "state 1" in the remaining. Both states are stable on the time scale of our experiments as long as non-invasive scan parameters are used ($|U| \leq 0.15$ V).

The topographic differences are accompanied by changes in the spectroscopic fingerprint of the molecules, as can be seen in the spectroscopy curves in Fig. 1f taken at the center of the MnPc molecules. MnPc molecules adsorbed on the flat $BiAg_2$ far away from an antiphase boundary, i.e., in the NOR configuration, show a single sharp peak at the Fermi energy (black curve), most likely an indication for a Kondo resonance. In contrast, the dI/dU spectra taken at the center of the deactivated molecule (state 0) on the antiphase boundary is featureless around Fermi energy, whereas a double peak structure is visible in the case of the activated molecule (state 1). The splitting of these peaks is further increased in the presence of a magnetic field (see chapter 1 of the Supplementary Material for details).

### Control and driving protocol for the switch

To determine the control parameters and the driving mechanism of the induced state switching we developed a measurement protocol,[13,14] some elements of which are sketched in yellow color in Fig. 1b and which is based on the following six steps (i)–(vi): first, (i) the STM tip is moved by a distance $d$ along the $[0\bar{1}1]$-direction away from the center of the MnPc molecule [yellow cross in Fig. 1b]. (ii) Then a bias pulse of $U_0$ is applied to set the initial state of the molecule in either state "0" or "1". Afterward, (iii) a topographic image of the molecule is measured with non-invasive scan parameters ($U = -0.1$ V; $I = 50$ pA) to confirm the initial molecule state. (iv) The tip is moved to the same position as in (i). (v) A bias pulse of a certain voltage $U_{Pulse}$ is applied for 4 s. Finally, (vi) the topography of the molecule is probed with non-invasive scan parameters again to analyze whether the molecule has switched to the other state. These six steps were repeated many times for statistical purposes.

By setting the initial pulse to $U_0 = 0.5$ V ($U_0 = -1.5$ V), an activation probability $p(0 \rightarrow 1)$ (deactivation probability $p(1 \rightarrow 0)$) is determined by dividing the number of activations (deactivation) by the total number of attempts, which is presented in Fig. 2a as a function of $U_{Pulse}$.

The activation starts for bias voltages lower than $U_{Pulse} < -0.8$ V and reaches saturation at $U_{Pulse} = -1.2$ V at a value of $p \approx 65\%$. We could not reliably probe large positive bias polarities, as voltages $U_{Pulse} > 0.8$ V very often lead to unwanted movements of the MnPc molecule. The thresholds for the deactivation are positioned symmetrically around the Fermi energy at bias voltages of $U_{Pulse} = \pm 0.3$ V. In the bias range of $0.4 V \leq |U_{Pulse}| \leq 0.8 V$ the deactivation probability is essentially 100%. When the bias is lowered below $U_{Pulse} < -0.8$ V it decreases. As indicated by the gray box in Fig. 2a there is some overlap of the energy ranges where both activation and deactivation processes can be observed, which was further analyzed by removing the first three steps (i)–(iii) of the measurement technique. In this way, the system was always excited with the same bias voltage $U_{Pulse}$ without resetting the state of the molecule to extract the probabilities for the four transitions [$p(1 \to 0)$, $p(0 \to 0)$, $p(1 \to 1)$, and $p(0 \to 1)$], which are shown in Fig. 2b. The data demonstrate, that the probabilities leading to the deactivated state [$p(1 \to 0)$, $p(0 \to 0)$] are decreasing, as the bias voltage is reduced from $U_{Pulse} = -0.8$ V to $U_{Pulse} = -1.3$ V. However, they never reach zero probability even at very negative bias voltages but maintain a value >20%, thereby hindering an activation probability of 100%.

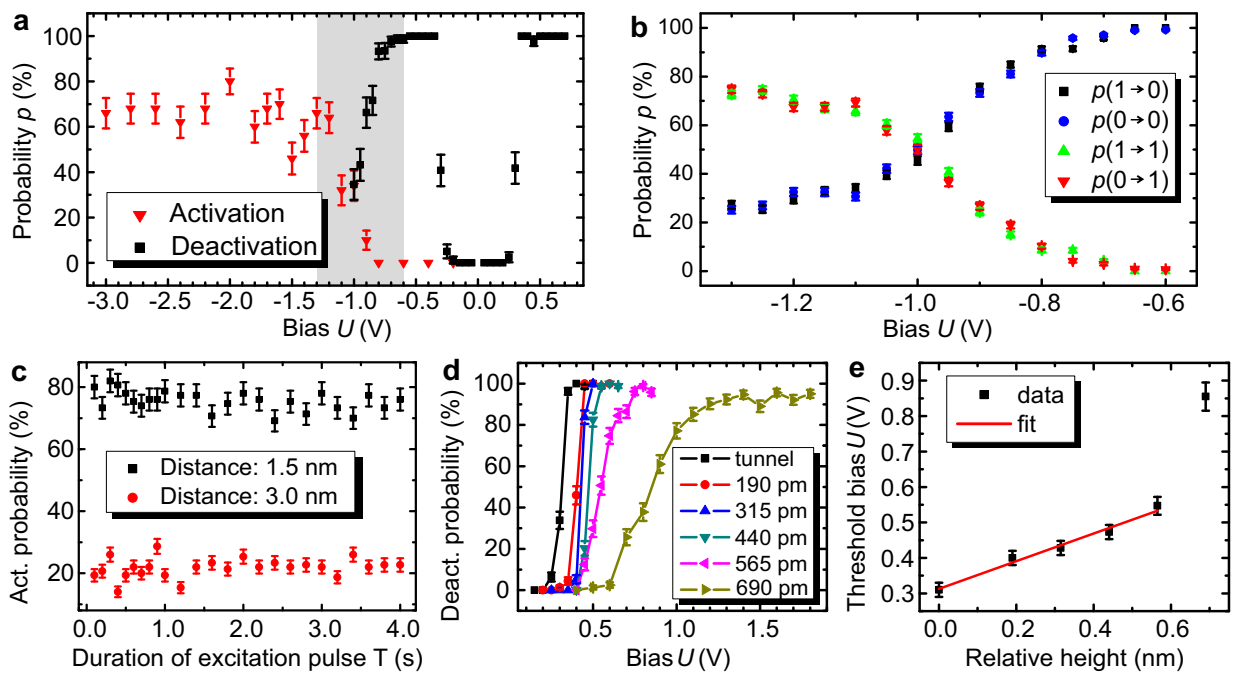

**Fig. 2** **a** Bias-dependent activation (red) and deactivation (black) probabilities taken at a lateral distance of $d = 1.5$ nm from the molecule [excitation parameters: $t = 4$ s; $I = 50/100$ pA (activation/deactivation)]. The measurement technique is described in the main text. **b** Bias dependency of the transition probabilities probed in the energy window marked in gray color in **a** (excitation parameters: $t = 4$ s; $I = 50$ pA). **c** Time dependency of the activation probability probed at two different distances (excitation parameters: $d = 1.5/3.0$ nm, $U = -1.5$ V, $I = 100$ pA). **d** Deactivation probabilities in dependency of different tip-sample distances $z$ (excitation parameters: $t = 4$ s; $d = 1.5$ nm.) The height reference point (0 pm = tunnel) was created with the scan parameter of $U = 0.35$ V and $I = 50$ pA with the tip being positioned at the excitation point ($d = 1.5$ nm). **e** Threshold bias voltages, which are defined as the points where the deactivation probabilities reaches a value of 50%, is plotted against the tip-sample distance $z$

Time-dependent measurements of the activation probability $p(0 \to 1)$ performed at two lateral injection distances $d = 1.5/3.0$ nm [cf. Fig. 2c] show that the activation probability is constant within the error bar, irrespective of the duration of the excitation pulse in the probed time interval from [0.1–4.0] s. We interpret these results as evidence that the time scale of the excitation pulses was much longer than the underlying switching processes. The driving mechanism of nonlocal excitations—with the tip being positioned away from the molecule—is usually based on either hot carriers[13,15–17] or on a field effect.[18–20] In the first case, charge carriers are injected in surface or bulk bands, diffuse to the molecule and transfer energy to the molecule, which leads to the excitation, whereas in the second case the process is induced by the electric field between tip and sample. The decreasing activation probability with increasing distance to the molecule (1.5 nm → 3.0 nm) would be in line with both scenarios, i.e., either a electric field effect, as the field is reduced with an enhanced tip molecule distance, or a hot carrier process. To analyze, which of these processes is responsible for the molecular switch, we modified step (iv) of the original measurement method [steps (i)–(vi)]. Additionally to a lateral shift of $d = 1.5$ nm, we also moved the STM tip a certain distance ($z$) vertically away from the substrate. The resulting data presented in Fig. 2d demonstrate, that deactivation is also possible if the tip is not in tunnel contact, which excludes a hot carrier process as the driving mechanism of the molecular switch. Furthermore, the threshold bias increases linearly with the vertical distance up to 565 pm, as can be seen in Fig. 2e. This trend is in line with the electric field approximation of a plate capacitor for the tip-sample junction.[21] If the tip-sample distance is further increased the approximation of a plate capacitor fails, which can be seen in the huge deviation for the last data point presented in Fig. 2e. The lateral distance dependency of the deactivation probability (cf. chapter 2 of the Supplementary Material) further confirms that the driving mechanism is an electric field effect. In total, the driving mechanism of the molecular switch is based on an electric field effect, which for low fields leads to the deactivation of the molecule and for higher fields to a competition between activation and deactivation.

Theoretical analysis
We have modeled the structure of MnPc on BiAg$_2$ on the antiphase boundary using the unit cell shown in Fig. 3 employing DFT + $U$ (see Methods for technical details). In all cases we used the experimentally determined adsorption site. Since the observed switching via an electric field is most probably concomitant with a change in the structure of the molecule, we focused on finding a structural bistability on the antiphase

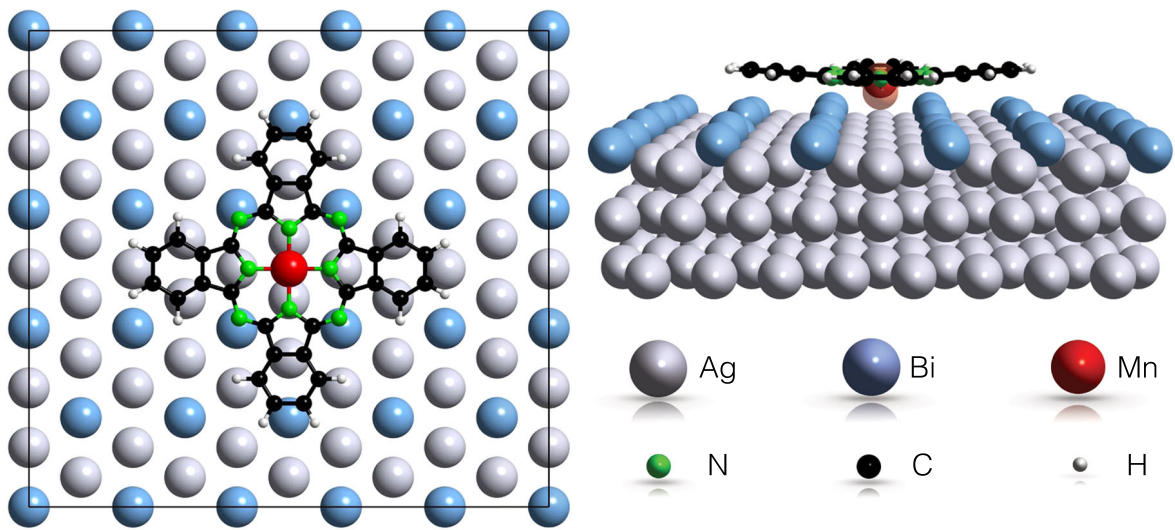

**Fig. 3** Top view of the unit cell employed in all calculations along with a side view of the fully relaxed structures showing the two possible positions of the Mn ion. In the side view we have superimposed the two structures (HI: Mn further away from substrate, LO: Mn closer to substrate), showing that the only qualitative change in the structure is the height of the Mn ion, the rest of the system remaining largely unperturbed

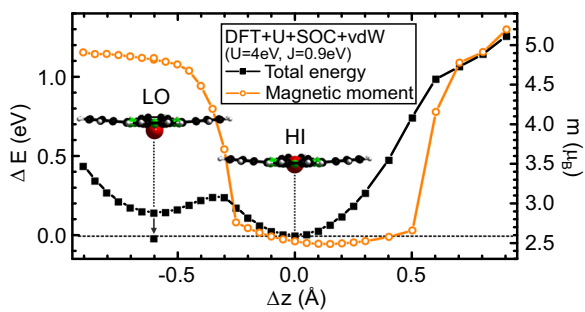

**Fig. 4** Scans of the total energy hypersurface on varying $\Delta z$. The calculations include all relevant interactions. The absolute value of the cell magnetic moment in z-direction (right ordinate) shows a minimum close to the fully relaxed structure at $\Delta z = 0$ and increases sharply when the Mn ion leaves the molecular plane. The values of $E$ and $m$ are also shown for the fully relaxed structure close to $\Delta z = -0.6$ Å. While the former reduces significantly, the latter remains very close to the unrelaxed value

boundary. Starting from a (flat) gas phase structure, the molecule relaxes into the geometry which we refer to in the following as HI, where the Mn atom remains in line with the molecular plane, see Figs. 3 and 4. In this geometry all arms of the molecule point slightly away from the surface like a cupped hand. Our first intuition was to expect that the phase boundary primarily modifies the geometry of the molecular arms. Therefore, we started by exploring various paths leading to a structure with all arms pointing towards the surface and also "saddle" structures with a reduced symmetry, where two arms (e.g., the ones aligned along the antiphase boundary) point towards and two arms point away from the surface. Varying the degree of bending of the molecule, as well as the adsorption height, however, did not lead to a stable second local minimum.

We therefore explored a shift of the central atom only, akin to SnPc.[5] In SnPc, due to the large size of the Sn ion, the central atom off-centers, thereby taking two symmetry-equivalent positions above or below the molecular plane. In our case a position outside of the molecular plane is only possible due to bonding to the edges of the antiphase boundary on the surface. By starting from the fully relaxed geometry in the HI position we changed the z coordinate of the Mn atom incrementally by up to $\Delta z = \pm 0.9$ Å. The rest of the atoms were kept fixed as a first step. Indeed reducing the height of the central Mn atom leads to a second local minimum in the total energy, as shown in Fig. 4 and reveals the local minimum at about $\Delta z = -0.6$ Å. This state, to which we will refer as LO in what follows, is stable against a full relaxation of molecule and surface alloy [cf. black arrow in Fig. 4]: $\Delta z$ only changes to $-0.58$ Å and the LO configuration becomes the ground state with $\Delta E = -17$ meV. Our calculations include local Coulomb repulsion within DFT + $U$, spin-orbit coupling ("SOC"), and van der Waals corrections ("vdW"). Crucial for the stability of the LO state is the existence of a magnetic moment [for further details, cf. chapter 4 of the Supplementary Material]. The energy differences involved are so small that we can for all intents and purposes call the structures energetically degenerate. The barrier between the two fully relaxed structures was estimated from a linear interpolation between the two to be about 500 meV. This constitutes an upper bound on the real barrier since the reaction path was not optimized e.g., using a nudged elastic band calculation.

At the level of DFT one difference between the two minima is the size of the magnetic moment. The LO structure is characterized by a very large moment of about $4.85\mu_B$, whereas the HI structure shows only $2.55\mu_B$. We observe a sharp change at $\Delta z = -0.3$ Å [see Fig. 4], brought about mostly by the Mn $d_{x^2-y^2}$. Details about the orbital occupation can be found in the chapter 7 of the Supplementary Material. We hence conclude that a sharp transition between the two spin states occurs concomitantly to a structural change between LO and HI configuration. A similar transition occurs for positive $\Delta z$, the structure in that case is however not stable. A way to stabilize it has been reported in ref.[22] by approaching the STM tip very close to the metal ion which in the case of FePc molecule on Au(111) was shown to cause a crossover from the Kondo to the spin-orbit regime.

Comparing these results to DFT calculations with MnPc placed on flat BiAg$_2$, i.e., in the NOR position, we conclude that the HI structure is very similar. This is corroborated by a marked similarity of the spin densities shown in Fig. 5a–c. On the other hand the LO structure is completely distinct from the other two. As there are no significant differences between the MnPc molecule in the HI state and that on flat BiAg$_2$—occupancies as well as spin densities—we infer that the HI state corresponds to state "1" observed by STM.

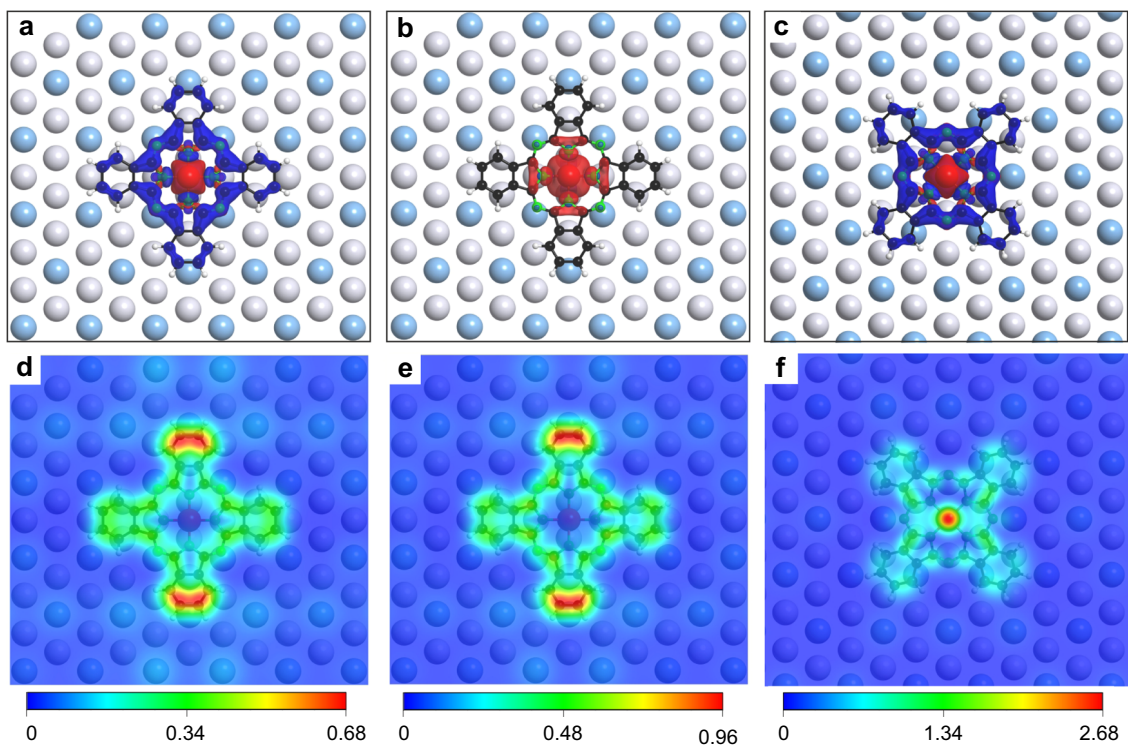

**Fig. 5** Spin densities (z component) of the HI state **a** the LO state **b**, as well as MnPc on flat BiAg$_2$ (NOR) **c** resulting from a DFT + $U$ + SOC + vdW calculation. Majority Spin shown in red, minority in blue. The isosurface value is the same for both spins and in every plot at 0.0025 $e$/Å$^3$. **d–f** Simulated STM image obtained by integrating the charge density resulting from a DFT + $U$ + SOC + vdW calculation between −0.1 eV and the Fermi level at 0 eV for the HI state **d** the LO state **e**, as well as NOR **f**. The charge density was sliced at 2 Å above the highest lying atom of the molecule and the colorbar is in units of 10$^{-9}$ $e$/Å$^3$

This assignment is further supported by an analysis of the height of the Mn atom for the activated molecule and that in the NOR case (cf. chapter 3 of the Supplementary Material). The topographic height difference of the Mn atom determined experimentally shows indeed only small deviation if we assume "1" to be the HI state and unrealistically strong deviations of about an order of magnitude assuming "1" to be the LO state.

In order to verify whether or not the charge distribution from DFT matches the experimentally observed topographic features, we evaluate the charge densities for the three different configurations in the energy range probed by the STM. The result is shown in Fig. 5d–f. Neither the LO nor the HI configuration displays an enhanced charge density at the central metal ion. On the contrary, our experimental results clearly demonstrate an accumulation of density of states for the activated molecule on the antiphase boundary, leading to a topographic protrusion at the Mn position. This indicates that the single particle description by DFT does not adequately account for this specific effect and many-body corrections need to be included.[8]

To this end we perform DFT + AIM calculations explicitly including the Coulomb interaction on the 3$d$ shell of Mn in a fully dynamical fashion (i.e., with frequency dependent self-energy), see Methods. The first important piece of information we obtain is that, by adding the many-body effects of the Coulomb interaction beyond the "DFT + $U$" level, the Mn 3$d$ shell in the LO state becomes much more correlated than in the other two cases, in particular way more than the HI configuration. This hints at LO being the non-Kondo molecular configuration, i.e., "0", in agreement with our previous conclusion based on the spin density distribution. The Kondo-active one, should indeed be characterized by the emergence of a local moment on the correlated $d$-shell, screened by the conduction electrons from the environment. This genuine many-body effect manifests itself in a Fermi-liquid like behavior of the self-energy which, at low-frequencies, becomes non-diverging, in contrast to the case of an impurity completely isolated from the surrounding (atomic limit). We can directly inspect the imaginary part of the self-energy on the Matsubara axis and easily rule out the Kondo effect in case of an atomic-like diverging self-energy.

We only focus on the orbitals that couple appreciably to the substrate, which in the present case, are the out-of-plane $d_{z^2}$ and the two other non-planar $d_{xz/yz}$ (see also ref.[8] and chapter 8 of the Supplementary Material). In the LO structure the self-energy of these orbitals is way bigger (cf. Fig. 6) than that of HI and NOR, up to an order of magnitude. The large self-energy of the LO configuration validates the assignment of LO being the deactivated molecule. In contrast, the HI and NOR configurations are characterized by much lower self-energies, indicating the presence of Kondo screening of the corresponding $d$-orbitals (see further discussion in chapter 9 of the Supplementary Material). Hence, the main outcome of our DFT + AIM calculations is that the self-energy is strikingly different between LO and HI, with the former being strongly atomic-like and the latter configuring itself as the Kondo-active molecule. These conclusions are also confirmed by an analysis of the orbital-resolved spectral functions, as shown in Fig. S7 of the Supplementary Material. As we more precisely discuss in chapter 9 of the Supplementary Material, it is well known that temperature effects are crucial in determining the shape of the spectral function of Kondo systems. The temperature of our quantum Monte-Carlo calculations, though being as low as 116 K, is still somewhat higher than the coherence scale of the $d_{z^2}$ orbital in the Kondo-active configuration (HI). In Fig. S7 of the Supplementary Material we indeed show that, while the spectra of $d_{xz}$ and $d_{yz}$ display a Kondo peak, the $d_{z^2}$ is characterized by a pseudogap and the corresponding self-energy is low but not yet in the perfectly linear regime characteristic of a Fermi-liquid.[23–25]

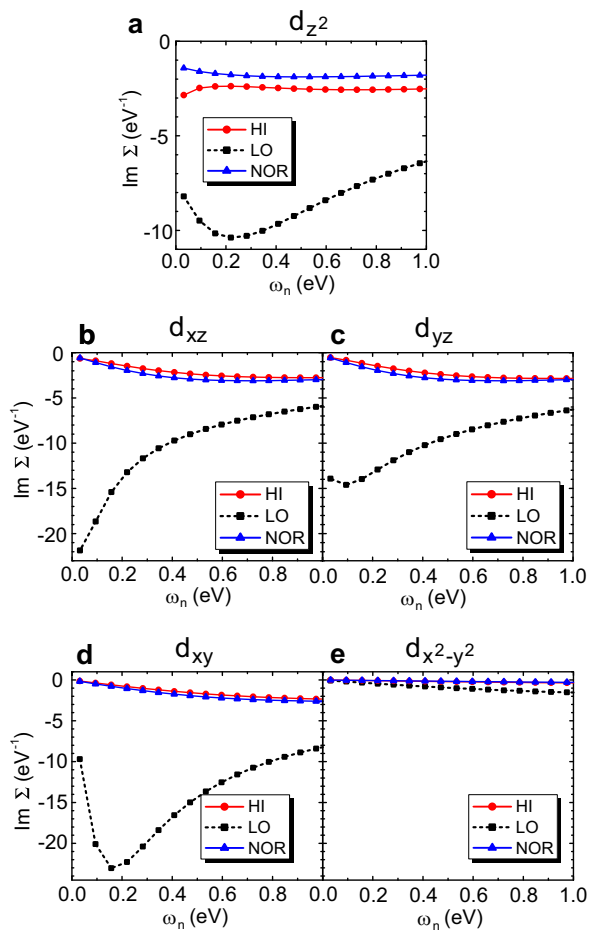

**Fig. 6** Matsubara self-energies at $T = 116$ K of the five orbitals $d_{z^2}$ **a**, $d_{xz/yz}$ **b** and **c**, $d_{xy}$ **d**, and $d_{x^2-y^2}$ **e** for the three structures considered. We have set the filling of the shell to 5 electrons. Further discussion about the analysis of the self-energies and their relation to the spectral functions can be found in chapter 9 of the Supplementary Material

In contrast, all orbitals display a strongly suppressed spectral weight in the LO configuration. For this reason, our global assigment and correspondence with experiments is expected to remain fully valid, even when future developments of the continuous-time quantum Monte-Carlo algorithms will allow us to reach even lower temperatures.

In conclusion, we designed a molecular spin switch by properly modifying the environment of a MnPc molecule, namely by positioning the molecule on top of an antiphase boundary of the BiAg$_2$ surface alloy. Due to bonding of the central metal ion to the edges of the boundary a second stable configuration is created with the manganese atom being moved out of the molecular plane. The electric field between the STM tip and the substrate can now be used to switch the molecule between the two states. Our results demonstrate that for the design of molecular electronics new paths can be opened by focusing not only on the molecule itself but also on its environment.

## METHODS
### Experimental
The Ag(111) crystal was prepared by cycles of 10 min Ar$^+$ sputtering with an ion energy of 1 keV and subsequent annealing up to 750 K for 10 min. After the last annealing cycle, the sample was held at a temperature of roughly 400 K and 1/3 of a monolayer of bismuth was evaporated on the substrate from a resistively heated crucible. After deposition, the sample was held for additional 15 min at the elevated temperature and afterwards cooled to room temperature. Finally, MnPc molecules (Strem Chemicals) were evaporated on the surface from a resistively heated crucible and the sample was directly transferred into either a home-built STM working at a temperature of roughly $T \approx 4.5$ K or a SPECS JT-STM with a base temperature of $T \approx 1.2$ K and a magnetic field of up to 3 T perpendicular to the sample surface. Constant current images were measured with the bias voltage $U$ attached to the sample. For scanning tunneling spectroscopy (STS) a small bias voltage $U_{mod}$ was added to $U$ and the $dI/dU$-signal was measured by detecting the first harmonic signal by means of a lock-in amplifier.

### Theoretical
We have modeled the structure of MnPc on BiAg$_2$ on the antiphase boundary using the unit cell shown in Fig. 3. The cell consists of three layers of Ag, where the topmost one forms the BiAg$_2$ surface alloy. In all cases we used the experimentally determined adsorption site. All crystal structures have been relaxed until the forces acting on each atom were smaller than 0.02 eV/Å using a $2 \times 2 \times 1$ mesh in reciprocal space. The relaxation involved the molecule as well as the BiAg$_2$ surface alloy. The DFT calculations were performed using the Projector-Augmented-Wave (PAW)[26] based Vienna ab initio simulation package (VASP).[27,28] The Perdew, Burke Ernzerhof version of the Generalized Gradient Approximation[29] was used in combination with the van der Waals dispersion correction due to Grimme (zero damping DFT-D3).[30] In selected cases, we performed comparisons with the superior optB88-vdW functional, finding the same qualitative trends. Since the Mn atom is close to a $d^5$ configuration the on-site Coulomb interaction is important (see Section "Results and Discussion") and was included in the calculations via DFT + U using the rotationally invariant formulation by Liechtenstein et al.[31] employing $U = 4.0$ and $J = 0.9$ eV. To account for the Bi atoms correctly we used the fully relativistic spin-orbit corrections in VASP.

The results obtained beyond DFT + U have been obtained constructing first a set of localized orbitals within the projection formalism (see refs.[32,33]) and then applying the numerically exact continuous-time quantum Monte-Carlo (CT-QMC) method[34] implemented within the hybridization expansion using the segment algorithm[35] (see ref.[36] for a review). More details on the CT-QMC calculations are given in chapter 10 of the Supplementary Materials.

### DATA AVAILABILITY
All relevant data are available from the authors upon request.

### ACKNOWLEDGEMENTS
This work was supported by the DFG through SFB1170 "Tocotronics" (project C02 and C07). M.K. and G.S. gratefully acknowledge the Gauss Centre for Supercomputing e.V. (www.gauss-centre.eu) for funding this project by providing computing time on the GCS Supercomputer SuperMUC at Leibniz Supercomputing Centre (www.lrz.de). D.S. acknowledges support from Spanish MINECO (MAT2016-78293-C6-6-R) and FEDER funds under the Interreg V-A program (POCTEFA 2014-2018, EFA194/16/TNSI)

### AUTHOR CONTRIBUTIONS
J.K. conceived the experiment. J.K. and A.K. carried out the STM experiments with support from D.S.; calculations were performed by M.K. with guidance from G.S.; and J.K., M.K., M.B., and G.S. participated in the analysis, figure planning, and draft preparation, which was completed with input from all the authors. The project was supervised by G.S.

### ADDITIONAL INFORMATION
**Supplementary information** accompanies the paper on the *npj Quantum Materials* website (https://doi.org/10.1038/s41535-018-0126-z).

**Competing interests:** The authors declare no competing interests.

**Publisher's note:** Springer Nature remains neutral with regard to jurisdictional claims in published maps and institutional affiliations.